# Optical properties of metal-dielectric-metal microcavities in the THz frequency range

Y. Todorov[1,*], L. Tosetto[1], J. Teissier[1], A. M. Andrews[2], P. Klang[2], R. Colombelli[3], I. Sagnes[4], G. Strasser[2] and C. Sirtori[1]

[1]*Laboratoire de Matériaux et Phénomènes Quantiques, Université Paris Diderot & CNRS - UMR 7162, 75205 Paris Cedex 13, France*
[2]*Solid State Electronics Institute TU Wien, Floragasse 7, A-1040 Vienna, Austria*
[3]*Institut d'Electronique Fondamentale, Université Paris Sud and CNRS, UMR 8622, 91405 Orsay, France*
[4]*CNRS, Laboratoire de Photonique et de Nanostructures (LPN), Route de Nozay, 91460 Marcoussis, France*
[*]*yanko.todorov@univ-paris-diderot.fr*

**Abstract:** We present an experimental and theoretical study of the optical properties of metal-dielectric-metal structures with patterned top metallic surfaces, in the THz frequency range. When the thickness of the dielectric slab is very small with respect to the wavelength, these structures are able to support strongly localized electromagnetic modes, concentrated in the subwavelength metal-metal regions. We provide a detailed analysis of the physical mechanisms which give rise to these photonic modes. Furthermore, our model quantitatively predicts the resonance positions and their coupling to free space photons. We demonstrate that these structures provide an efficient and controllable way to convert the energy of far field propagating waves into near field energy.

## 1. Introduction

One of the most intensively studied topics in modern photonics is the ability to control the light at sub-wavelength scales. Because of their extremely high electron concentration, metals have unique optical properties which make them suitable for this objective. In particular, metal-dielectric surfaces are able to support surface plasmon-polariton modes which - at frequencies comparable to a fraction of the metal plasma frequency - can be tightly localised around the metallic surface [1,2]. This is the basis of many novel photonic applications in the visible and near-infrared [3,4]. At longer wavelengths, surface plasmon waveguides had a real impact on the quantum cascade lasers [5,6], where the size of a dielectric waveguide would be incompatible with standard epitaxial techniques.

However, the sole use of surface plasmons for building compact practical devices could become inadequate for longer wavelength range, since the surface plasmon extension into the dielectric greatly increases as the metal loss is decreased for frequencies far below the plasma frequency [1]. Then the metal structuring comes to the rescue. For instance, spoof surface plasmons in a structured metallic membrane have been recently proposed [7] and demonstrated in the THz region [8]. Most generally, metal surfaces which are a patterned on a sub-wavelength scale can be engineered to sustain resonant modes and then new properties emerge. Metamaterials are recent prominent example [9,10].

On the other hand, in the far infrared range of the electromagnetic spectrum metals are not only low loss, but also excellent reflectors. Precisely this property can be exploited to confine the electromagnetic field far beyond the diffraction limit. It is well known that the fundamental TM mode guided between two metallic plates does not have a cut-off frequency, and the light can be squeezed into sizes much smaller than the wavelength [11]. Structuring one of the metal layers can provides an efficient way to couple in and out the radiation, which is important for devices such as THz lasers in a distributed feedback configuration [12,13] or photonic crystal lasers [14,15]. The ability of metal-metal structures to support strongly localised electromagnetic fields has already been pointed out in experiments performed in the microwave part of the spectrum [16,17].

In this paper we systematically explore, both experimentally and theoretically, the properties of metal-semiconductor-metal structures in the THz frequency range with subwavelength dimensions. The top metal layer is periodically patterned, which leads to the formation of resonant modes with an electromagnetic field that is squeezed into the thin semiconductor layer. Such device architecture is particularly appealing since the semiconductor layer can naturally be filled with – for instance - a quantum engineered active medium.

The paper is organised as follows. In part 2 we study the simplest periodic metallic pattering: the rectangular strip grating. In 2.1 we will report on our experimental studies in the THz frequency range, and compare them with a full numerical model [18]. In 2.2 we will study the electromagnetic field of the resonant modes and we will show how their symmetry affects the coupling efficiency of the structure. In 2.3 we analyze in details the confining mechanism trough a simpler and more intuitive analytical approach. In 2.4 we study the influence of the geometrical parameters of the structure on the dispersion properties of the photonic modes. In part 3, we present our experimental results with bi-dimensional gratings, which are analyzed on the basis of the concepts developed in part 2. In part 4 we demonstrate the important role played by the evanescent near field of the grating for the resonant coupling of free space photons into the highly confined electromagnetic modes. Finally, in the last part of the manuscript we will return to the 1D geometry to show a peculiar topological property, which yields two different types of resonances respectively in the THz and MIR frequencies, yet supported by the same structure.

The basis of our full numerical model is the so called "modal method", introduced in 1981 by Botten et al. [19] and Scheng et al. [20]. This model has been extensively applied to study free standing lamellar metallic grating and the photonic phenomena associated. The implementation of the modal method that we use is described in [18]. Metal losses have been taken into account by the surface impedance boundary condition, which is an excellent approximation in the THz and MIR frequency range [21].

## 2. 1D sub-wavelength grating

*2.1 Experimental study*

In Figure 1(a), we illustrate the typical structure of our investigation: a thin semiconductor (Gallium Arsenide) slab of thickness $L$, sandwiched between a metallic (gold) mirror and a lamellar grating of strip width $s$ and pitch $a$. The periodicity of the grating is $p=a+s$. The inset of Fig. 1(a) indicates the typical dimensions of the grating region, which is in the order of tens of microns. Note that for the frequency region of interest 1-9 THz ($\lambda$=300-34µm) the typical grating periods remain smaller than the wavelength $\lambda$. The thickness of the gold stripe is $h$=450nm.

The optical response of the structure is tested, in the THz region, using reflectometric measurements. In our setup a polarized beam from a Globar lamp of a Bruker interferometer (IFS66) is focused on the grating and the reflected intensity is measured with a cooled Bolometer detector. The incident angle $\theta$ indicated in Fig. 1(a) is either 10° or 45°, depending on the experimental configuration. While the plane of incidence in the experiments is always perpendicular to the grating stripes, we will also discuss a more general case. The incident light is *p*-polarized, i.e. with an electric field perpendicular to the stripes. In the orthogonal *o*-polarization the reflectivity of the structure is close to unity in the studied frequency range, therefore we do not consider this polarization [18].

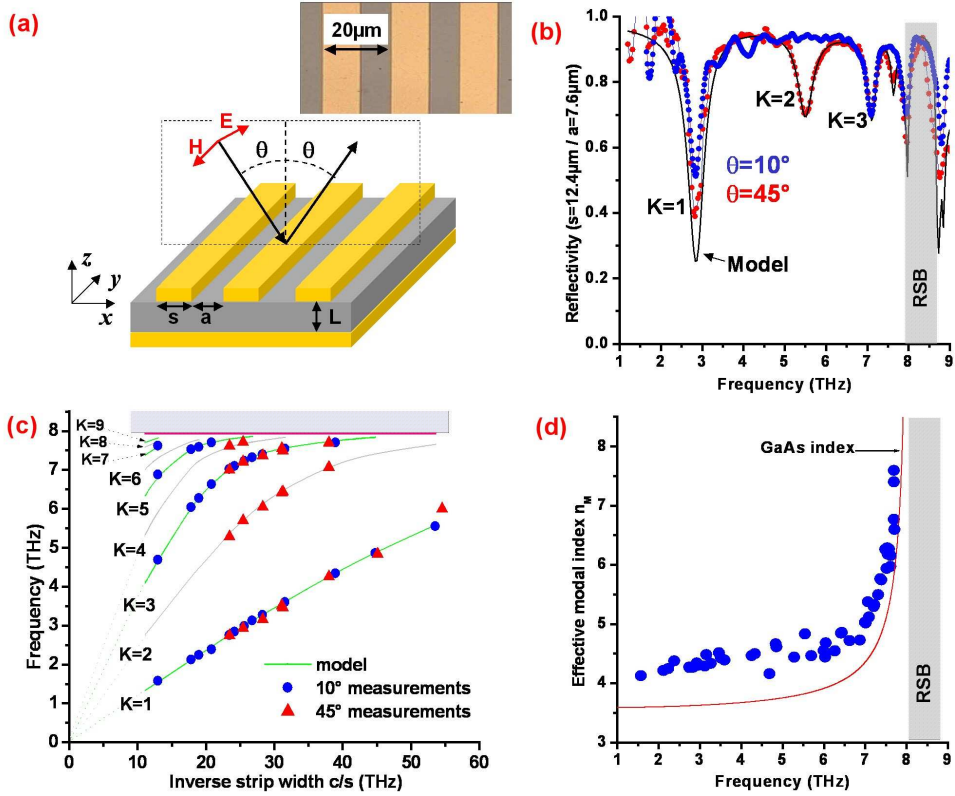

Fig.1. (a) Schematics of the structure. A semiconductor slab with thickness $L$ is sandwiched between a metallic plane and a lamellar metallic grating. The inset shows a top view of a sample with typical dimensions. The direction and the polarization of the incident plane wave is also indicated. (b) Experimental spectra for incident angles $\theta=10°$ and $45°$ (dotted curves) and simulation at $\theta=45°$ (continuous line). (c) Resonant frequencies of the reflectivity dips as a function of the inverse strip width $c/s$. (d) Effective index of all the resonances as deduced from Eq. (2) (dots). The red continuous line is the bulk semiconductor index $n_s$. For all plots, the restrhalen band of GaAs is indicated in grey.

In Figure 1(b) we provide two typical reflectivity spectra of a structure with $s=12.4$ µm and $a=7.6$ µm, for both incident angles $10°$ and $45°$. The grating period is $p=a+s = 20$ µm. The thickness of the semiconductor slab is $L=0.8$µm. In this limit no diffraction is possible and the reflectivity signal is collected from the $0^{th}$-order specular reflection only. As no energy can be diffracted into high-order free space modes, the reflectivity dips observable in the spectra correspond to resonant absorption within the structure. We can actually speak about resonant photon tunnelling [22], a phenomenon appearing whenever the frequency of the incident photon matches the frequency of a photonic mode of the structure. When this condition is satisfied incident photons are coupled into the mode and their energy is eventually dissipated by inherent ohmic losses of the metallic walls [23], yielding the absorption minima in the spectra. The mechanism of resonant absorption will be studied in details in paragraph 4.

The absorption features are labelled by an integer $K=1, 2, 3…$. From the data of Fig. 1(b) one can see that the odd $K$ resonances are excited at both angles of incidence, whereas the even $K$ resonances are absent at $10°$ (which is almost normal incidence). This indicates that the coupling with the resonant modes of the structures obeys precise selection rules that will be detailed in paragraph 2.2.

A recurrent feature in the experimental spectra is visible in the band between 8 and 9 THz. This frequency range corresponds to the well known *restrhalenband* region, where the electromagnetic radiation resonantly interacts with the optical phonon of polar semiconductors [24].

In Fig. 1(b) we have indicated, as a black continuous line, the prediction of the full numerical model which takes into account the dispersion of the semiconductor refractive index and the ohmic losses of the metal trough a Drude-like dielectric function for the metal layers:

$$\varepsilon(\nu) = 1 - \frac{1.52 \times 10^{30}}{\nu(\nu - i1.6 \times 10^{13})}, \quad (1)$$

where the frequency $\nu$ is expressed in Hz. Using Eq. (1) numerical computations precisely account for the positions and the widths of the resonances (for simplicity, only the simulation for $\theta = 45°$ is presented).

To characterize the absorption resonances, we have studied gratings with different stripe widths and periods, at both angles of incidence: 10° and 45°. A summary of the data and comparison with the model are presented in Fig. 1(c), where the peak positions are plotted versus the inverse of the stripwidth $1/s$. In order to express both axes in units of frequency, we rather use $c/s$ where $c$ is the speed of light. It can be seen from the figure that the frequency positions are independent from the angle of incidence. However, for $\theta = 10°$ only the odd $K$ resonances are observable. For frequencies that are sufficiently far from the *restrahlenband* the peaks positions evolve almost linearly with $c/s$, and we can write the simple expression [25]:

$$\nu_K = \frac{K}{2n_M} \frac{c}{s} \quad (2)$$

This expression corresponds to a standing wave under the grating stripes, with the integer $K$ counting the electric/magnetic field nodes/maxima [25] (see also Figure 2). In the phenomenological expression Eq. (2) we have introduced the effective modal index $n_M$ of the standing wave. Knowing the geometry of our structures and the resonant frequencies the effective index can be easily computed from Eq. (2), and has been plotted in figure Fig.1(d), for all orders $K$. We observe that the behaviour of $n_M$ is related to the dispersion of the GaAs material index (in red) [26]. Both $n_M$ and the refractive index of the semiconductor rapidly increase while approaching the restrhalen band, causing the saturation of the dispersion curves $\nu_K(c/s)$ of Fig. 1(c). However, the value of $n_M$ is ~14% higher than the bulk index. This difference arises from the impedance mismatch that confines the electromagnetic field under the stripes, and yields a reflection phase shift at the edges of the resonator that is not an exact multiple of $\pi$. This effect is studied in details in paragraph 2.3.

*2.2 Field maps and selection rules*

In Figure 2 we plot the electromagnetic field at the resonant frequencies as calculated by the full numerical model used to fit the experimental spectra of Fig1(b), For *p*-polarized light the only tree non-zero field components are $E_z$, $H_y$ and $E_x$, with a coordinate system $Oxyz$ oriented as in Figure 1(a). The origin of the system is taken on the left metal stripe, and the plane $z = 0$ corresponds to the upper semiconductor surface. In Figure 2(a) these components are plotted for the first excited resonance $K=1$, under normal incidence $\theta = 0°$.

A standing wave pattern in the *x*-direction is clearly visible in the contour-plots of $E_z$ and $H_y$. These components are essentially confined below the grating stripes, and in these regions they can be approximated by the expressions:

$$E_z(x) \approx E_z \cos\left(\frac{\pi K}{s} x\right), \quad H_y(x) \approx H_{y0} \sin\left(\frac{\pi K}{s} x\right), \quad x \in (0, s). \quad (3)$$

These components are therefore independent from *z*. On the contrary, the $E_x$ component is strongly localized around the metallic corners. This already has been noticed for similar structures in the microwave region [16] and can be identified with the following phenomenological expression, based on the field map of Figure 2(a):

$$E_x(x, z=0) \propto E_z(x=0, z=0)\delta(x) - E_z(x=s, z=0)\delta(x-s), \quad (4)$$

with $\delta(x)$ the Dirac delta function. This expression is defined here in an intuitive way. However, it can be rigorously justified as it will be shown in paragraph 2.3.

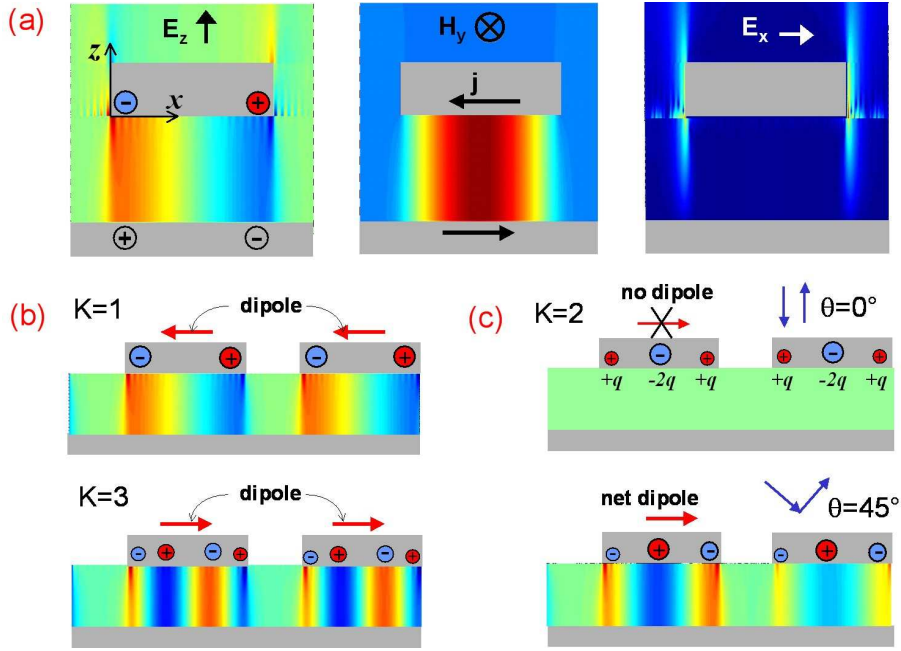

Fig. 2. (a) Plot of the tree components ($E_z$, $H_y$ and $E_x$) of the electromagnetic field for the first ($K=1$) resonance, under normal incidence $\theta=0°$. The maximal values are in red and minimal in blue. The induced charges and currents are obtained trough Eq. (9). (b) Electric field $E_z$ for the first odd resonances ($K=1,3$). The arrow indicates the dipole moment of the induced charge distribution on the metallic stripe. (c) Expected charge distribution of the $K=2$ mode at normal incidence $\theta=0°$, and net induced dipole at oblique incidence.

The expression of the *x*-component of the electric field Eq.(4) can be used to establish the selection rules of the resonances. To this end, we compare the latter with the *x*-component of the diffracted field in the air. We recall that the diffracted field can be cast into a Raleigh decomposition of evanescent and propagating components [27]. In the case of grating with sub-wavelength period *p*, as in our case, the only propagated component is the 0$^{th}$ diffraction order, which is a superposition of the incident wave and reflected wave with amplitude $R_0$:

$$E_x^0(x, z>0) = (R_0 e^{i\gamma_0 z} - e^{-i\gamma_0 z})e^{i\alpha_0 x}, \qquad (5)$$

$$\alpha_0 = k_0 \sin\theta = \frac{2\pi}{\lambda}\sin\theta, \qquad \gamma_0 = \sqrt{k_0^2 - \alpha_0^2}. \qquad (6)$$

Here $k_0=2\pi/\lambda=2\pi\nu/c$, $\alpha_0$ and $\gamma_0$ are the *x*- and *z*- wavevector components of the incident plane wave. We have dropped for the moment the evanescent components of the diffracted field as they will be discussed in details in paragraph 4. Neglecting the thickness *h* of the metallic grating, which is much smaller than the wavelength, we can match the tangential electric fields Eqs. (4) and (5) at *z*=0. Integrating the continuity condition over the grating period we obtain:

$$R_0 - 1 = \frac{1}{p}\int_0^p e^{-i\alpha_0 x} E_x(x, z=0^-) dx. \qquad (7)$$

This integral is proportional to the quantity $I_K$:

$$I_K(\theta) = 1 - (-1)^K e^{i\frac{2\pi s}{\lambda_K}\sin\theta} = 1 - (-1)^K e^{i\frac{\pi K}{n_M}\sin\theta}. \qquad (8)$$

Here we have used Eqs. (2-4), and $\lambda_K=c/\nu_K=2n_M s/K$ is the wavelength of the $K^{th}$ resonance. The depth of the resonances in the reflectivity spectra (see Fig.1(b)), which we can define as $1-|R_0|^2$ is proportional to the quantity Re($I_K$) which therefore contains the selection rules stated above. Indeed, according to (8) the odd *K* orders are excited independently from the incident angle, since $(-1)^K=-1$ and $I_K(\theta)$ is never zero. Instead, the even *K* orders are strictly prohibited at normal incidence $\theta = 0°$, and become deeper at oblique incidence following an approximate $\sin^2\theta$ dependence.

An intuitive picture of the selection rules can be grasped by examining the distribution of charges and currents induced on the metallic walls, which are provided by:

$$\sigma = \mathbf{E}.\mathbf{n}, \quad \mathbf{j} = \mathbf{H} \times \mathbf{n}. \tag{9}$$

with σ the sheet charge density, **j** the surface current density, and **n** the outgoing normal to the metallic walls. The surface currents and charges induced by the mode have been indicated in Fig. 2(a). Since $E_z$ and $H_y$ are *z*-independent, from Eq.(9) we deduce that the induced charges and currents in the lower metallic plane are the reverse images of the ones induced on the stripes. Therefore the structure naturally has a capacitor and an inductive behaviour. For instance, for the *K=1* mode there are two capacitors which are located at the edges of the stripe, while the inductance is located in the centre.

On the other hand, it is clear from Eq. (3) that for *K=1* the electric field $E_z$ changes sign from one strip edge to the other, and it is maximum in absolute value around the edges. Therefore from Eq. (9) the charges induced on the metal stripe are located around the edges and have opposite signs (the latter is also imposed by the overall neutrality of the charge distribution). The charge distribution on the stripe has, therefore, a net *dipole moment* in the *x*-direction. This dipole moment is parallel to the electric field of the *p*-polarized incoming wave, which for $\theta = 0°$ also oscillates in the *x*-direction, thus the coupling between the mode and the incident radiation is allowed. This is actually true for any angle of incidence $\theta$, since the incoming electric field will always have a non zero projection on the dipole moment of the charge distribution. The same reasoning can be applied also for the *K=3* resonances, illustrated in Fig. 2(b), and in general any odd resonance.

The same conclusions will be obtained if we consider the surface currents rather than the surface charges. Indeed, it is clear from Fig. 2(b) that, because of the induced surface currents, the structure also possesses a non-zero *magnetic moment* that is parallel to the incoming magnetic field. Note that in the semiconductor region the current loops are completed by the displacement current $\varepsilon_0 \partial E_z / \partial t$. It can be readily shown that odd resonances will also have non-zero net magnetic moment that always allows the coupling. In other words one can say that each stripe of the grating behaves like a microwave patch antenna [28].

Now let us examine the possible dipole moment of the *K=2* mode, under normal incidence. The electric field mode, as seen from Eq. (3), has two maxima at the metallic edges and a minimum in the middle of the metal stripe. The expected induced charge density is indicated on figure 2(c). Under normal incidence, it is symmetric with respect to the middle of the stripes and consists of charges +*q* at the strip edges and -*2q* in the middle. This can be viewed as two opposite dipoles oscillating in counter phase, with no net dipole. The *K=2* and all the even modes are therefore strictly forbidden under normal incidence.

On the contrary the coupling is allowed when the symmetry is broken under oblique incidence. In this case, the incident electromagnetic field excites the opposite dipoles with a delay proportional to $\sin(\theta)$. The charges oscillate slightly out of phase and create a small net dipole, as indicated in Fig. 2(c).

*2.3 Impedance mismatch*

We turn now to the discussion of the confining mechanism which is responsible for the formation of the standing wave patterns described above. This mechanism arises from the impedance mismatch between the metal-metal and single metal regions [25]. Indeed, when the thickness *L* is much smaller than the wavelength λ, the metal-metal region supports only the lowest order guided $TM_0$ mode, which does not have a cut-off frequency. On the contrary, in the open single metal regions the electromagnetic field can be seen as a continuum of plane waves (Fig. 3(a)). The modal mismatch of the electromagnetic field at the openings of the double metal regions causes the reflection of the $TM_0$ at the openings, as illustrated in Fig. 3(a), hence the formation of the standing wave pattern. A similar phenomenon is found at the open end of acoustic instruments [29]. In this paragraph we propose a simple model to evaluate the impedance mismatch and we provide an analytical formula for the resonance frequencies $\nu_K$.

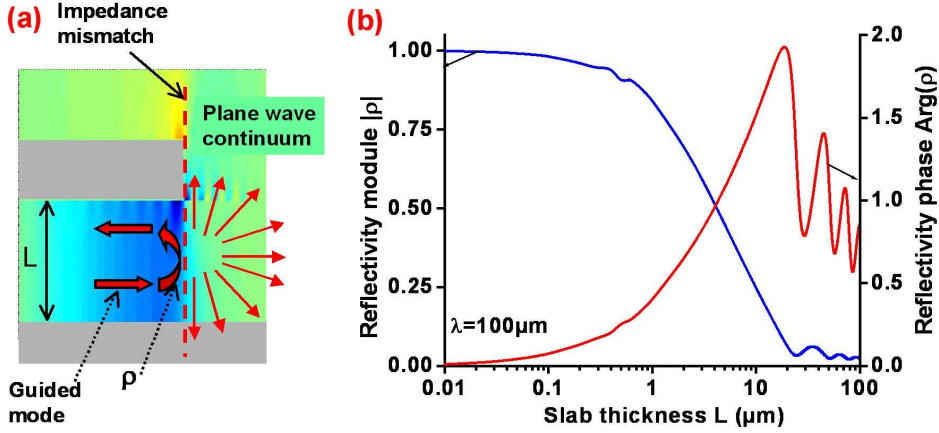

Fig. 3. (a) Illustration of the impedance mismatch between the $TM_0$ guided mode $L$ and the radiation continuum in the single metal region. (b) Plot of the module and the argument of the complex reflectivity $\rho$ (Eq.(13)), as a function of the thickness $L$ for a wavelength $\lambda=100\mu m$ (frequency $\nu=3$ THz ), in the single-mode approximation.

The field of the metal-metal regions between the stripe and the metal mirror can be written as a superposition of two counter-propagating guided $TM_0$ modes:

$$(-L \leq z \leq 0; \; -\infty < x < 0), \quad E_z^{TM} = A\exp(i\beta_x x) + B\exp(-i\beta_x x), \qquad (10)$$

with $\beta_x$ the propagating constant of the $TM_0$ mode and $A$ and $B$ the amplitudes of the counterpropagating fields. The propagation constant is $\beta_x = n_g k_0$ where $n_g$ is the guided mode effective index which can be obtained from the usual guided mode theory. It is actually slightly higher than the bulk semiconductor index represented in figure 1(d) due to the penetration of the mode in the metallic regions.

On the contrary, the semiconductor slab of the single metal region cannot support any guided modes. To model the impedance mismatch, we will assume that this region is a semi-infinite space entirely composed of semiconductor, supporting a continuous set of plane waves:

$$(-L \leq z \leq \infty, \; 0 < x < \infty), \quad E_z(x,z) = \int_0^{+\infty} C(k_z)\exp(ik_x x + ik_z z)dk_z, \quad k_x = \sqrt{n_s^2 k_0^2 - k_z^2}. \quad (11)$$

Here $n_s$ is the bulk semiconductor index, and for simplicity, we consider only one open end, at $x>0$, as in Fig 3(a). Then the effective reflectivity coefficient which describes the impedance mismatch is defined as:

$$\rho = B/A. \qquad (12)$$

In order to carry out the calculation for $\rho$, we change our structure in order to obtain a system that has essentially the same physical properties, but that can be resolved analytically. First, the lower limit of the integral of Eq. (11) is pushed to $-\infty$ in order to obtain an invertible Fourier transform. Then all the metal regions are replaced by semi-infinite thick metal walls, in order to take advantage of the electric field boundary condition $E_z=0$. By doing this transformation we obtain a single slit perforated in a metallic membrane of width $s$. This geometry has been treated in the literature [30, 31]. Here we provide the final result:

$$\rho = \frac{n_s - \tilde{n}_g}{n_s + \tilde{n}_g}, \quad \tilde{n}_g = n_g \frac{2\pi}{\varphi} \frac{1}{\int_{-\infty}^{+\infty} \frac{\text{sinc}^2(t\varphi/2)}{\sqrt{1-t^2}} dt}. \quad (13)$$

Here $\varphi = n_s k_0 L$. In figure 3(b) we plot complex reflectivity $\rho$ as a function of the cavity thickness $L$, for a wavelength $\lambda = 100\mu m$ ($\nu = 3$THz). The plot clearly shows that the impedance mismatch becomes more and more important as the slab thickness $L$ is decreased as compared to the wavelength, and it vanishes for thick resonators [32].

Once the effective reflectivity $\rho$ at the stripe edges is known, we can compute the frequencies and the effective index of the standing waves. Since the metal stripe can be considered as a short waveguide of length $s$, in order to have resonant modes, the following condition must be satisfied:

$$1 - \rho^2 \exp(2i\beta_x s) = 0. \quad (14)$$

This condition states that the field under the stripe should recover its phase after a roundtrip between the two resonator open ends. Since both the effective reflectivity $\rho$ (Eq.(13)) and the propagation constant $\beta_x$ are complex functions of the frequency $\nu$, then in general Eq. (14) must be solved numerically and would provide complex solutions for the frequencies of the resonant modes. However, our goal is to avoid the numerical computation and to provide handy analytical expressions with clear physical meaning. A first step is to neglect the imaginary part of the guided mode index $n_g$, and then, having in mind that $\beta_x = 2\pi \nu n_g/c$, we express the complex solutions of Eq. (14) as:

$$\tilde{\nu}_K = \frac{c}{2sn_g}\left(K - \frac{\text{Arg}(\rho)}{\pi}\right) + i\frac{c}{2sn_g}\frac{\ln|\rho|}{\pi}. \quad (15)$$

The real part $\text{Re}(\tilde{\nu}_K)$ of Eq.(15) provides the frequencies of the standing waves. It can be directly related to the experimental data through comparison with Eq. (2), defining an apparent mode index $n_M$ as:

$$n_M = \frac{n_g}{1 - \frac{\text{Arg}(\rho)}{\pi K}}. \quad (16)$$

The imaginary part of Eq. (15) provides the damping rate, and hence an estimation of the quality factor of the confined modes. In this model, the damping comes from both the metal loss (contained in the guided mode index $n_g$) and the radiation into the continuum of plane waves. The quality factor that we obtain is:

$$Q_K = \frac{\text{Re}(\tilde{\nu}_K)}{-2\text{Im}(\tilde{\nu}_K)} = \frac{\pi K - \text{Arg}(\rho)}{-2\ln|\rho|}. \quad (17)$$

In order to compare Eqs.(16) and (17) with the experiment, we take the values of $\rho(\nu)$ and $n_g(\nu)$ at the measured experimental frequencies $\nu = \nu_K$. The results of this study are shown in figures 4(a) and 4(b). The experimental values of the quality factors $Q$ have been obtained through lorenzian fits of the experimental spectra. Knowing the physical dimensions $s$ of the resonators, a fairy good agreement is obtained for the effective modal index $n_M$ (Fig. 4(a)). This validates our estimation of the reflectivity phase shift due to the coupling with the continuum of radiated waves.

The above simplified model does not account properly for the quality factors in this first order approximation (Fig. 4(b)), but still it provides the correct order of magnitude. On the contrary, the quality factor is very well reproduced using the numerical modal method. We believe that this discrepancy arises partly from the periodicity of the grating which is not taken into account in the estimate of the impedance mismatch. The periodic arrangement of the resonators modulates the plane wave continuum, which changes the way that the resonator energy is lost through dissipation and radiation.

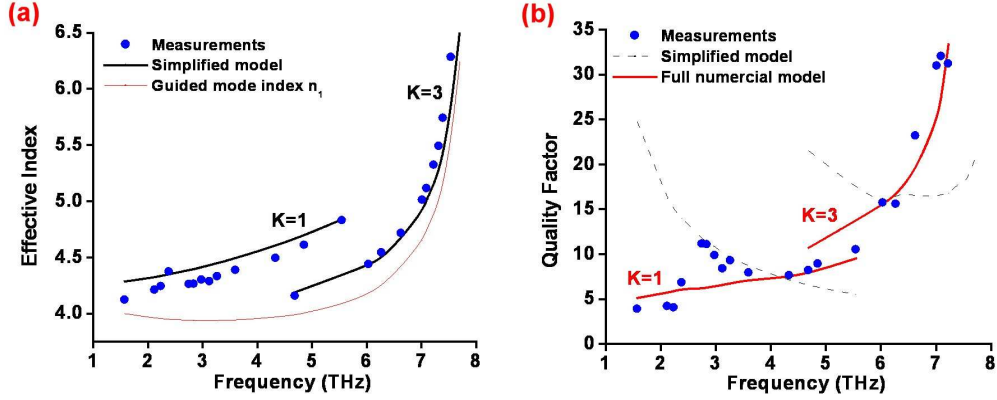

Fig. 4. (a) Effective index of the *K=1* and *K=3* resonances (dots) compared with theoretical predictions from Eq. (16) (black lines). The effective index $n_g$ of TM$_0$ mode of a planar double-metal waveguide of thickness $L=0.8\mu m$ is also indicated. (b) Quality factors as deduced from experiment (dots), full model (red continuous line) and the simplified model from Eq. (17) (dashed curve).

Finally, we will show that the analytical model described here also accounts of the strong $E_x$ field singularities around the metallic corners described in 2.2. Sparing the reader some lengthy calculations, the function $C(k_z)$ from Eq.(11) is evaluated to be:

$$C(k_z) = \frac{E_z(x=0^-)}{2\pi i} \frac{1-e^{-ik_z L}}{k_z \sqrt{n_s^2 k_0^2 - k_z^2}}. \tag{18}$$

Similar result can be found in [30]. Then using Eq. (11) and the Maxwell equation $\partial_x E_x + \partial_z E_z = 0$ the field $E_x$ component around the corner of the metallic stripe can be expressed as:

$$E_x(x,z) = E_z(x=0^-) \frac{1}{2\pi i} \int_{-\infty}^{+\infty} \frac{e^{ik_z z + i\sqrt{n_s^2 k_0^2 - k_z^2} x}}{\sqrt{n_s^2 k_0^2 - k_z^2}} dk_z. \tag{19}$$

This integral describes a strongly localized function around the metallic corner ($x = 0$, $z = 0$). For instance, at *z=0* we obtain $E_x(x, z=0) \propto H_0^1(2\pi x n/\lambda)$, with $H_0^1(x)$ the 0$^{th}$ order Hankel function of the first kind, which is singular at *x=0*, whereas the integral clearly has a delta-like behaviour in the *z*-direction [33].

Physically, Eq. (19) describes a cylindrical wave with a source located at the metal corner of the stripe, with an amplitude that is proportional to the vertical component of the electric field at the resonator end $E_z(x=0^-)$. This vision confirms our description of the coupling selection rules in 2.2. Indeed, when the incident plane wave strikes the structure it excites two secondary sources with amplitudes $E_z(x=0)$ and $E_z(x=s)$, located at the resonator ends. These secondary sources are excited with a phase delay provided by the finite optical path of the incident wave along the metallic stripe (see also Eq. (8)). Low reflectivity is obtained when the two interfere destructively, which is always true for odd *K* resonances and is impossible for resonances with even *K* under normal incidence.

*2.4. Dispersion properties*

*2.4.1 Dispersion along the grating*

The microcavity configuration described above, producing localized modes, is different from the one usually found in the literature [34], where the periodic structures rather yield delocalized photonic crystal modes, with defined dispersion relations. This regime can also be recovered in our structures as hinted by Figure 3(b) which shows that the reflectivity $\rho$ vanishes for thick resonators (large $L$). Therefore, by increasing the thickness of the structure, the localized modes couple with each others and become de-localized along the whole grating. This is best understood by studying the dispersion properties of the system, as illustrated in figure 5. In this figure, the real part of the eigenfrequencies of the system are computed in a scattering-matrix approach and plotted in the first Brillouin zone. The x-axis of these plots is the normalized parallel wavevector $k_{ZB}=k_0 \sin\theta \; (p/\pi)$. We recall that $\theta$ is the incident angle from the air side and $p$ is the grating period.

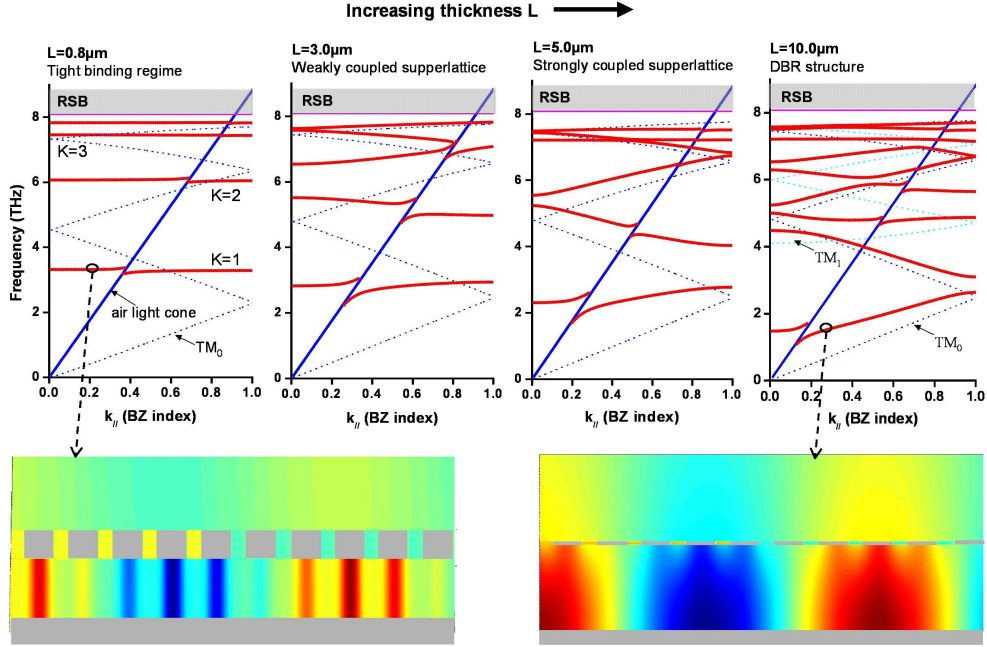

Fig. 5. Dispersion diagrams for structures with increasing slab thickness $L$ and identical gratings ($s$=11.5µm, $a$=5.5µm) . The blue continuous line is the air light cone. The blue dotted lines are the folded dispersion of the guided TM modes of a planar double metal waveguide with the same thickness as the grating structures. The two insets show the magnetic field distributions $H_y$ for structures with $L$=0.8µm and $L$=10µm (not at the same scale).

For a very sub-wavelength thickness $L$=0.8µm an essentially flat dispersion is obtained, which indicates localized modes. Indeed, in this "tight binding" regime, the different resonators are uncoupled and their frequency depends only on the stripe width $s$ (Eq. (2)), and depend neither on the incident angle, nor on the period $p$. As illustrated by the contour-plot of the magnetic field in figure 5, the field is a superposition of the field distributions of the individual localized resonators, modulated by a Bloch-Floquet envelope $\exp(ik_{ZB}\pi x/p)$. In this case the periodicity creates a set of infinitely degenerate photonic modes, each labelled by the index $k_{ZB}$. In the same figure we also plot the dispersion of the $TM_0$ mode folded into the first Brillouin zone to show the deviation of the localised modes with respect to the case of the planar dispersion.

When the structure thickness is gradually increased, the different resonators start to couple, resulting in a more dispersive behaviour. The anti-crossing feature that appears as the dispersion encounters the air light-cone is related to the Wood anomalies [35,36]. Finally, for a very thick structure $L$=10 µm, we recover the usual DBR-like behaviour. The field distribution in this case is dominated by the Bloch-Floquet envelope. Far from the Brillouin zone edges the photonic dispersion is parallel to the folded double-metal planar waveguide modes $TM_0$ and $TM_1$, whereas the grating opens gaps in the Brillouin zone edges, which are proportional to the coupling constant of the DBR [37]. Therefore the group velocity of the photonic mode, that is proportional to $d\omega/dk_{//}$, can be changed from 0 to a maximum value $c/n_s$ by controlling the semiconductor thickness of our structure.

Similar behaviour is obtained for a fixed thickness, when the stripe separation $a$ is reduced. The minimal value of $a$ that couples the resonators depends strongly on the thickness $L$ and the metallic loss. This value is roughly provided by the decay length of the field outside the resonator can be estimated at $\lambda \, \text{Im}(\tilde{n}_g)/n_g$, where $\tilde{n}_g$ is defined in Eq. (13). It can be deduced from Eq. (13) and the reflectivity plot of Fig. 3(b) that this decay length strongly decreases as the cavity thickness is reduced. For instance, at $L = 0.8\mu m$, the modes start to couple for $a < 0.5 \, \mu m$. In order to scale the reflectivity resonances to higher frequencies, while keeping them uncoupled, one must not only decrease the size of $s$, but also decrease the thickness $L$ and/or increase the stripe separation $a$.

*2.4.2 Dispersion along the slits*

So far, we have been concerned with the dispersion properties of the structure in the direction of the grating periodicity. It is clear that, because of the 1D nature of the metallic texture, the subwavelength modes described above are strictly dispersionless only along this direction. Considering now an incident plane that is *parallel* to the stripes, with an incident wavevector $k_y$ it is easy to show, using the Helmholtz equation, that the formula providing the resonant frequencies is:

$$\nu^2 = \left(\frac{cK}{2n_M s}\right)^2 + \frac{c^2}{4\pi^2 n_s^2}k_y^2. \qquad (20)$$

The resonances are still excited in *p*-polarization only, because only this polarization provides an electric field with non-zero projection perpendicular to the metal stripes. In practice the dispersion described by (20) is difficult to observe experimentally due to the high refractive index $n_s$ of the dielectric (GaAs) used in our structures. Introducing $\phi$ as the oblique angle of incidence in the air, we can write $k_y = \nu \sin\phi/2\pi c$ leading to:

$$\nu^2 = \left(\frac{Kc}{2n_M s}\right)^2 \frac{n_s^2}{n_s^2 - \sin^2 \phi}. \qquad (21)$$

Since the semiconductor dielectric constant $n_s^2 \gg 1 > \sin^2\phi$, the $\phi$ dependence in Eq. (21) could be neglected, as we have also verified by numerical simulations.

**3. 2D structures**

In order to obtain strictly dispersionless resonances the metal layer has to be structured in two dimensions. The simplest example is the square patch grating described in figure 6. The side of the patch will be labelled $s$, and the period of the square grating is labelled $p$. Fig. 6(a) describes an experiment with a grating with dimensions $s = 13\mu m$ and $p = 17\mu m$, where the probing wave arrives at an incident angle of 45° with respect to the normal of the grating surface. The incident direction is parallel to the patch edges (along the X-Γ direction, in crystallographic conventions). The structure is probed both in *p*- and *o*- polarizations, with the electric field parallel (*p*-) /perpendicular (*o*-) to the plane of incidence. With this 2D surface patterning the absorption features are observed for both polarizations (Fig. 6(b)). In analogy with the 1D case, these resonant features can be identified with TM$_{NM}$ modes confined under the metallic patches, with frequencies provided by the equation:

$$\nu_{NM} = \frac{c}{2n_M s}\sqrt{N^2 + M^2}, \qquad (22)$$

and the electric field distribution under the patches is approximately given by:

$$E_z(x, y) \approx E_{z0} \cos\left(\frac{\pi N}{s}x\right)\cos\left(\frac{\pi M}{s}y\right), \qquad (23)$$

that is, again, independent from $z$. The effective index $n_M$ takes approximately the same values as for the 1D grating. The major difference arises from the fact that now the field is confined in the tree dimensions of space and therefore the resonance frequencies are exclusively determined by the resonator size, *regardless the angle incidence*. In this respect, these resonances are the substantiation of truly dispersionless photonic modes [17].

Our aim is now to understand the complex structure of the experimental spectra of figure 6(b). We first remark that, according to Eq. (22) the modes $(K,0)$ and $(0,K)$ are degenerate, and according to Eq.(23) their electric field oscillates in two perpendicular directions (along the *x*- and *y*- direction, respectively). Therefore these resonances can be excited independently from the polarization, and even by unpolarized light, since any polarization can be decomposed into two orthogonal linearly polarized components. This is indeed the case in the experiment, but only for even orders $K$, whereas the odd orders are excited only in *p*- polarization. Further, the (1,1) resonance is not very well pronounced.

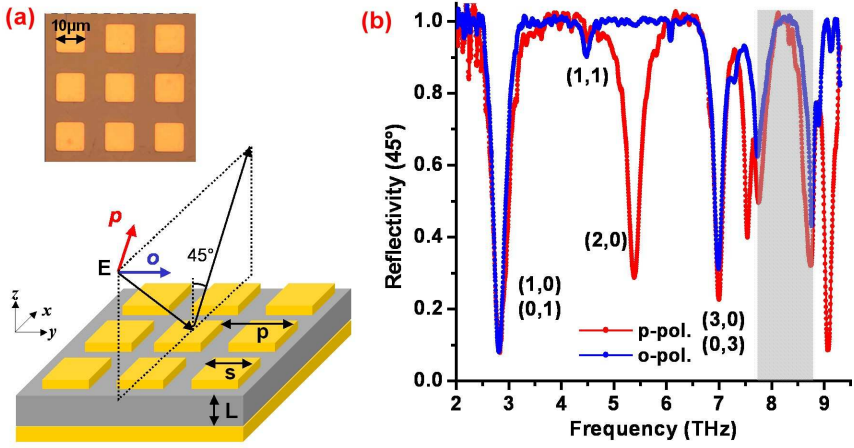

Fig. 6. (a) Structure with a square patch grating. The inset shows a top view of a sample with typical dimensions. The incident direction and the possible polarizations of the incident beam are also shown. (b) Reflectivity spectra of a sample with dimensions $L$=1.5µm, $p$=17µm and $s$=14.3µm for both *o*- and *p*- polarized incident beam. The incident angle is $\theta$=45° with respect to the grating normal.

As for the 1D case, we are interested in defining the reflection selection rules. These rules can be formulated quantitatively in the same way as in the 1D case, with the use of the in-plane electric field component $\mathbf{E}_{//}$. We first introduce the in-plane wavevector component $\mathbf{k}_{//}$ of the incident wave, that can be expressed through the polar incident angles ($\theta, \phi$):

$$\mathbf{k}_{//} = (k_0 \sin\theta \cos\phi, \ k_0 \sin\theta \sin\phi), \qquad (24)$$

The experiment corresponds to $\theta$=45°, $\phi$=0° and both polarisations. Then the superposition of the incident and reflected 0$^{th}$ diffracted order is:

$$(p\text{-polarization}) \qquad \mathbf{E}_{//}^{00} = \frac{\mathbf{k}_{//}\gamma}{k_0^2}\left(R_{00}e^{i\gamma z} - e^{-i\gamma z}\right)e^{i\mathbf{k}_{//}\cdot\mathbf{r}}$$

$$(o\text{-polarization}) \qquad \mathbf{E}_{//}^{00} = \frac{(\mathbf{k}_{//}\wedge\mathbf{u}_{//})\gamma}{k_0^2}\left(R_{00}e^{i\gamma z} - e^{-i\gamma z}\right)e^{i\mathbf{k}_{//}\cdot\mathbf{r}} \qquad (25)$$

Here $\mathbf{r}=(x,y)$, $\mathbf{u}_{//}=(\mathbf{u}_x,\mathbf{u}_y)$ are the in-plane unit vectors, and $\gamma=\sqrt{k_0^2-\mathbf{k}_{//}^2}$. Using a delta-like form for the modal electric field which is similar to the 1D case, we have:

$$\mathbf{E}_{//}(x,y,z=0) \propto \left[-E_z(0,y)\delta(x+s)+E_z(s,y)\delta(x)\right]\mathbf{u}_x \\ +\left[-E_z(x,0)\delta(y+s)+E_z(x,s)\delta(y)\right]\mathbf{u}_y. \qquad (26)$$

Here the coordinates $x$ and $y$ run only along the metallic edges on the resonator. Following 2.2 we have to match the in-plane component of the field. However, to take into account of the vector character of the field, we rather match the projections $\mathbf{k}_{//}.\mathbf{E}_{//}$ for p- and $\mathbf{k}_{//} \wedge \mathbf{E}_{//}$ for o-polarization. In this way, the modal field is projected only in the direction of the incident field. As a result, we obtain the two functions $I_p$ and $I_o$ which describe the relative intensity of the resonances:

$$I_p \propto \oiint_{\substack{resonator \\ edge}} \frac{\mathbf{k}_{//}.\mathbf{E}_{//}}{\mathbf{k}_{//}^2} e^{-i\mathbf{k}_{//}\mathbf{r}} dxdy, \quad I_o \propto \oiint_{\substack{resonator \\ edge}} \frac{\mathbf{k}_{//} \wedge \mathbf{E}_{//}}{\mathbf{k}_{//}^2} e^{-i\mathbf{k}_{//}\mathbf{r}} dxdy. \quad (27)$$

These integrals run over contour of the resonator. Dropping the constant field amplitudes, we obtain the following result for the special case ($\phi=0°$) of figure 6(a):

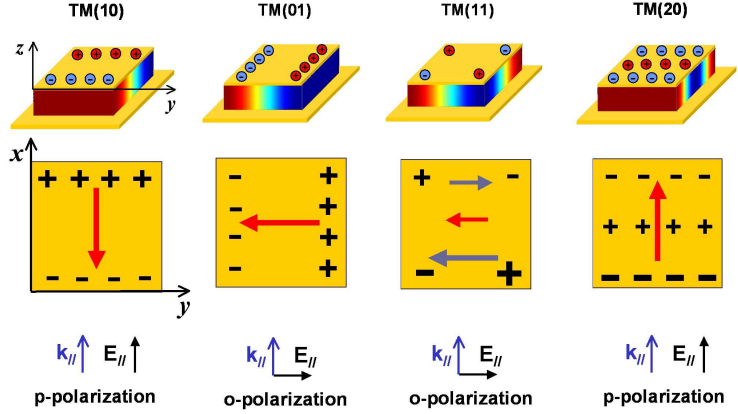

Fig. 7. Electric field $E_z$ distribution (Eq. (23)) and induced charge distribution (through Eq.(9)) for the first excited modes. Here $\mathbf{k}_{//}$ and $\mathbf{E}_{//}$ are the projections of, respectively, the wavevector and the electric field of the incident wave on the grating surface (xy-plane). For each mode, we have indicated with red arrows the net dipole induced under these experimental conditions.

$$I_p = (1-(-1)^N e^{i\Omega_{NM}})\delta_{0M}, \quad I_o = (1-(-1)^M)\frac{1-(-1)^N e^{i\Omega_{NM}}}{\pi^2 N^2 - \Omega_{NM}^2}\Omega_{NM}. \quad (28)$$

where $\Omega_{NM} = 2\pi s \sin\theta / \lambda_{NM}$ and $\lambda_{NM}$ is the wavelength of the resonance (N,M). It is easily verified that the above expressions predict correctly the observed selection rules, and qualitatively the relative depth of the resonances, which is proportional to Re($I_{p,o}$). Note that this method can be easily generalized for any incident direction, and for an arbitrary combination of linear polarizations as well as resonators of different forms.

Again, an intuitive picture of these results can be obtained invoking the induced charges on the edges and the retardation effects, as it is illustrated in figure 7, where we have indicated $\mathbf{k}_{//}$ and $\mathbf{E}_{//}$ for each experimental configuration. For even K resonances, the induced edge charges always have opposite signs, and there is a non-zero dipole moment, independently on the retardation effects. Therefore these resonances will always be excited. On the contrary, the (1,1) resonance has a quadrupolar distribution of charges. This can be viewed as a two opposite edge dipoles. When the two dipoles are misbalanced because of the retardation effects, as in figure 7, then a small net dipole moment appears which is perpendicular to the propagation of the incident field. This explains why these resonances are observable only in o-polarization, and with very small intensity (Fig. 6(b)). Similarly, the (2,0) resonance can be observed only in p-polarization, because the misbalanced charge distribution creates net dipole only along the direction of propagation $\mathbf{k}_{//}$.

## 4. Resonant absorption

The formalism developed in paragraphs 2.2 and 3 allows one to determine in a straightforward way the reflectivity selection rules that determine which resonances will be excited and which not, for a given experimental configuration. These selection rules depend solely on the geometrical form and the size of the metallic stripe or patch. However, they do not provide the *absolute* values of the resonance depths, which quantify the resonant absorption. Indeed, once an incident photon is resonantly captured into the strongly localized photonic modes, its destiny will be governed by a balance between the loss rate on the metallic walls and the rate at which it is in/out-coupled. This balance depends on the properties of the metal, as well as on the other geometrical parameters of the structure, namely the grating period $p$ and the semiconductor slab thickness $L$.

The $0^{th}$ order incident and reflected propagating waves are sufficient to obtain the resonator form factors Eq. (8) and Eq. (28). To go further we need to take into account the evanescent waves that are created in the diffraction process [27] and which are tightly confined around the grating surface. The aim of this section is to elucidate the role played by the near field of the grating for the energy loss of the photonics modes on the metal walls, which sets the absolute value of the reflectivity at resonances $R_{min}$. As we have already mentioned in paragraph 2, the resonance depth 1-$R_{min}$ describes the ability of the structure to capture the energy of the incident wave, that is, the *coupling efficiency* of the photonic resonances.

Experimentally, the quantity 1-$R_{min}$ may be a subject to uncertainties due to the baseline of the spectra. In order to minimise these uncertainties, we take a more precise definition of the resonant absorption lineshape. This is obtained noting that the photonic resonance corresponds to a complex pole $\nu_K + i\Delta\nu$ of the scattering matrix [38]. Therefore, if we denote by $r$ the amplitude of the reflected beam around the resonance, from general considerations based on the scattering matrix approach [22] it can be cast in the form $r(\nu) = 1 - iA/(\nu - \nu_K + i\Delta\nu)$. This leads to a Lorentzian lineshape of the energy reflection coefficient $R(\nu) = |r(\nu)|^2$:

$$R(\nu) = 1 - \frac{1 - R_{min}}{1 + (\nu - \nu_K)^2 / \Delta\nu^2}, \qquad (29)$$

and now the coupling efficiency 1-$R_{min}$ can be extracted from a Lorentzian fit of the absorption lineshape.

In order to understand the grating effects on the resonant absorption, we have studied systematically the evolution of the absorption lineshapes as a function of the grating period $p$ and structure thickness $L$, both for 2D and 1D structures. In figure 8(a) and 8(b) we report the results obtained with the fundamental patch modes of the 2D gratings, in the experimental conditions described in the previous paragraph. Very similar results have been obtained also for 1D gratings, which is expected since 2D and 1D structures have identical symmetry along the direction of periodicity. In figure 8(a) we compare the spectra of two structures with identical gratings ($p = 17.0\mu m$ and $s = 10.4\mu m$), but two different values of the semiconductor slab thickness $L = 0.85\mu m$ and $L = 1.5\mu m$. The experimental spectra are very well fitted by Lorentzian lineshapes, which allows us to extract the coupling efficiency 1-$R_{min}$ according to Eq. (29). A coupling efficiency of almost 100% is obtained for the first mode of the thicker ($L = 1.5\mu m$) resonator. In figure 8(b) we plot the coupling efficiency 1-$R_{min}$ as a function of the strip width $s$ for different periods $p$ and for both values of $L$. The values of the coupling efficiency are independent from the patch width $s$, but instead depend strongly on the grating period. This is clearly seen for instance in figure 8(b) in the proximity of $s=13.0$ μm where the efficiency abruptly changes by changing the grating period. **A general conclusion from the data is that shorter periods have better coupling efficiencies.** Moreover, all periods of the $L = 1.5\mu m$ structure have very high coupling efficiencies, and absorb almost all of incident light at resonance, at least for the subwavelength values of $p$ explored in the experiment.

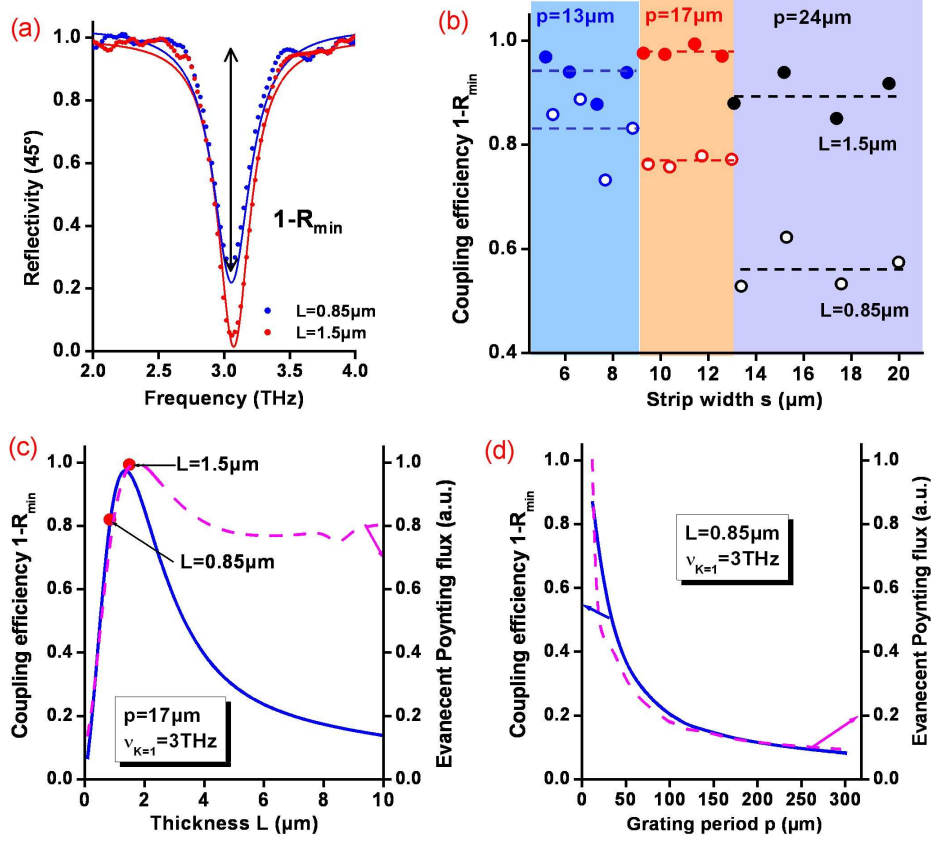

Fig. 8. (a) Two reflectivity spectra (dotted curves) for two different values of the slab thickness ($L$=0.85µm and $L$=1.5µm) and the corresponding lorenzian fits (continuous lines). The grating parameters are $p$=17µm and $s$=13.0µm. The coupling efficiency $1$-$R_{min}$ is defined as the amplitude of the lorenzians profile Eq. (29). (b) Experimental values of the coupling efficiency as a function of the strip widths $s$, for various parameters $L$ and $p$. The dashed lines are guide for the eye. (c) Simulated coupling efficiency and Poynting flux (Eq. (33)) as a function of the slab thickness, for a grating with $p$=17µm and $s$=13.0µm. The dots indicate the experimental values. (d) Simulated coupling efficiency and Poynting flux as a function of the grating period, for a slab thickness $L$=0.85µm and $s$=13.0µm.

The fact that the energy transfer is favoured for shorter grating periods is already an indication that the evanescent field plays an important role. For simplicity, we will illustrate this assertion in the 1D case. At the end, the conclusions that will be drawn will be independent from the particular geometry of the grating. Anticipating that the energy transfer will take place mostly in the semiconductor slab, where the electromagnetic field of the modes is predominantly located, we write now the full expansion of the $H_y$ field in into Rayleigh-Bloch harmonics in that region:

$$H_y = \sum_n e^{i\alpha_n x}(P_n e^{i\gamma_n^S z} + Q_n e^{-i\gamma_n^S z}), \quad Q_n = r_n P_n e^{2i\gamma_n^S L}, \tag{30}$$

$$\alpha_n = k_0 \sin\theta + \frac{2\pi}{p}n, \quad \gamma_n^S = \sqrt{n_s^2 k_0^2 - \alpha_n^2}. \tag{31}$$

Here $n$ is an integer and $r_n$ is the reflectivity on the planar metal mirror, which is well-known function of the diffracted order $n$. The expansion Eq.(30) contains two sets of slab waves counter-propagating in the $z$-direction, with amplitudes $P_n$ and $Q_n$. These amplitudes can be obtained exactly from the numerical model in Ref. [18]. To grasp the general trends of the dependencies of $P_n$ and $Q_n$ on the expansion order $n$, we can also use the Fourier-transform of Eq. (3) which provides:

$$P_n, Q_n \sim \int_0^s H_{y0} \sin\left(\frac{\pi K}{s}x\right) e^{-i\alpha_n x} dx \sim \frac{\sin(\alpha_n s/2)}{\alpha_n^2 s^2/4 - 1}. \tag{32}$$

Here, for simplicity, we have the dropped constant and phase prefactors. For sufficiently high values of n Eq.(32) can be approximated by sinc($\pi n s/p$) where "sinc" denotes the cardinal sine function $\sin(x)/x$.

In the expansion Eqs.(30) and (31) the evanescent waves correspond to the terms that propagate along the grating with in-plane wavevectors $\alpha_n$ greater than that of light $n_s k_0$ in the substrate and decay exponentially over a distance $1/\operatorname{Im}\gamma_n^S$ from the grating or the metal plane.

From the definition (31) it is clear that the condition $\alpha_n > n_s k_0$ will involve more expansion orders with non-vanishing amplitudes (Eq.(32)) if the period $p$ is short. For the case of the subwavelength gratings ($p<\lambda$) in our experiments the modal fields (Eq. (3)) decompose essentially on the evanescent waves of the grating, since only the $0^{th}$ order of the decomposition of Eq.(30) is a propagating wave.

The role of the evanescent waves for the resonant absorption is revealed trough their contribution to the Poynting flux in $z$-direction $S_z = 1/2 \operatorname{Re}(E_x H_y^*)$ (the star indicates complex conjugate). Although a single evanescent wave which is exponentially decaying in the $z$-direction can not carry a Poynting flux in that direction, a *pair* evanescent waves decaying in opposite directions will have a non-zero Poynting flux. This effect is, for instance, at the origin of the quantum-mechanical tunnelling across a finite potential barrier in quantum mechanics [39], or the attenuated internal reflection in optics [40]. In our case, a non-zero Poynting flux arises because of the interference between the evanescent pairs in the expansion of Eq.(30). The tunnelling Poynting flux of the slab evanescent waves, averaged over the grating period is then:

$$S_z = \sum_{n(ev)} \frac{|P_n|^2}{\gamma_n^S} e^{-2\gamma_n^S L} \operatorname{Im} r_n. \qquad (33)$$

Here the sum runs on the evanescent orders only. As expected, the flux Eq.(33) vanishes if the metal mirror is to be removed ($r_n=0$), or if it were a perfect conductor ($\operatorname{Im}(r_n)=0$).

We now compare the quantity of Eq.(33) as well as the coupling efficiency $1-R_{min}$ by numerical simulations for various parameters $L$ and $p$. For consistency with the experiment, the incident angle is 45° angle of incidence, and we use the first $K=1$ resonance with $\nu_{K=1} = 3$ THz. Two types of simulations have been performed. First, in figure 8(c) the grating period $p=17.0$ µm is fixed, and we vary the cavity thickness $L$. In the second, shown in figure 8(d), the thickness $L$ is fixed to a value $L=0.85$µm, and the grating period is spanned to very high values. For all plots, the coupling efficiency $1-R_{min}$ is plotted in blue, and the Poynting flux is in magenta and, for clarity, it has been normalized to its maximal value.

From figure 8(c) we observe that the coupling efficiency reaches 100% around a value of $L=1.5$µm, but decays rapidly for very thin resonators. The experimental data from fig 8(a) are also indicated in red circles. Similar behaviour is recovered with the Poynting flux (Eq.(32)), and the two curves coincide for thin resonators. We stress that this thickness range is particularly pertinent, since for higher thicknesses the lateral confinement of the resonances is lost (see paragraphs 2.3 and 2.4). The correlation between the two curves therefore attests the role of the evanescent waves to mediate the energy loss to the metallic walls. Moreover, these results show that we are able to push our system into an optimal regime of coupling, which is similar to the *critical coupling* regime between a waveguide and a resonator [22, 41]. In this regime, the resonator loss (in our case, the loss on the metal walls) equilibrates the capture rate for the incoming photons, and all of the incoming energy is absorbed by the resonator. Moreover, in this regime, the far field incoming energy is entirely converted into energy of the near field.

The strong correlation between the coupling efficiency and the evanescent Poynting flux is further confirmed in the second simulation, in figure 8(d). Indeed, both are strongly decreasing with the increasing grating period. This strong decrease is explained by the fact that, according to Eq.(31) and Eq.(32), as the grating period $p$ is increased, less and less evanescent waves with non-vanishing amplitudes are available for the energy transfer.

The fact that the energy transfer is mediated through the evanescent waves has two important consequences. First, for short enough periods, the resonant absorption is almost independent from the incident direction. This is clearly seen from the experimental data of figure 1(b), and can be understood from the definition of the wavevector $\alpha_n$ (Eq.(31)). Indeed, for strongly subwavelength gratings $p \ll \lambda$ the term $2\pi n/p$ dominates the angle dependent term $2\pi \sin\theta/\lambda$, and the amplitudes (Eq.(32)) of evanescent waves are insensitive to the incident direction. Second, if the semiconductor slab has some intrinsic absorption e.g. due to electronic transitions, the evanescent field would help the energy transfer to the electronic transition in the dielectric as well as to the metal. Therefore such devices could be very useful for realising angle independent detectors.

## 5. MIR resonances: Back to the 1D case

Unlike the 2D square patch structure, the 1D grating is actually subject to a topological frustration: should it be considered as a grating of slits, or a grating of stripes? So far, we have pointed out the *p*-polarized modes observable in the THz region, which are strongly localized under the grating stripes. However, because of this topological ambiguity, under certain conditions, the open slits regions could also sustain confined modes. Namely, for the same grating structure, such modes are observable for *o*-polarization and for shorter wavelengths, in the mid-infrared spectral region. The experimental setup is identical as described in previous section, by now the far-infrared bolometer detector is replaced by Mercury Cadmium Telluride (MCT) detector operating in the $\lambda=3\mu m-15\mu m$ range.

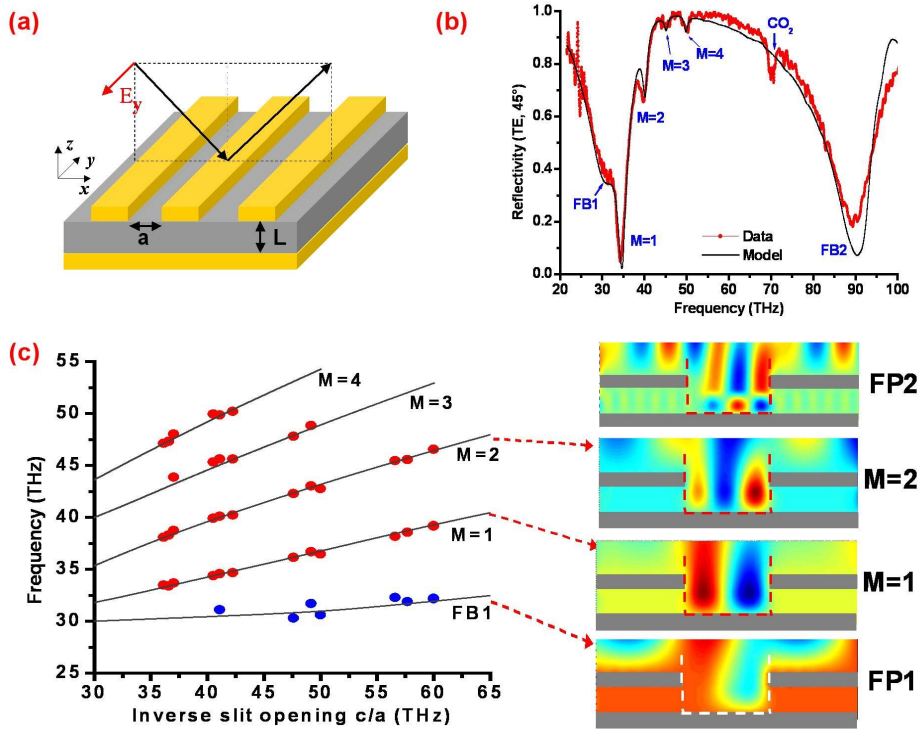

Fig. 9. (a) Experimental configuration for probing the structure reflectivity in the mid-infrared spectral range. (b) Typical reflectivity spectra for a structure with parameters $p=20\mu m$ and $a=7.3\mu m$. The $CO_2$ atmospheric absorption is also indicated. (c) Dispersion diagram for the observed resonances, obtained from multiple reflection measurements. The dots are experimental measurements and the continuous lines are numerical computation. The electric field $E_y$ for various mode orders is also plotted.

These resonances are shown in figure 9 where we present the experimental configuration in (9(a)) and a typical spectrum (9(b)). The measured reflectivity of the *o*-polarized beam exhibits two broad features and a series of equally spaced sharper resonances. The experimental spectrum is very well reproduced by our numerical model [18]. The analysis of many spectra, that is summarized in figure 9(c) shows that the resonance position are independent from the period $p$ or the strip width $s$, but shift almost linearly with the inverse of the slit opening $a$. The nature of these features becomes clear while examining the electric field distribution $E_y$ (Fig. 9(c)). We recall that this is the only non-zero component of the electric field for the $o$-polarization. The field contour plots show that indeed the electric field is localized in the single metal regions, and now vibrates both in horizontal ($x$) and vertical ($z$) direction. The field distribution is therefore governed by two integers $M$ and $K>0$. The first counts the number of field maxima ($M+1$) in the lateral direction, whereas $K$ counts the maxima in the vertical direction.

These variations translate the boundary conditions imposed on the frontiers of the region indicated in red or white dashed line on the contour plots. The vertical contribution arises from the fact that the electric field has to vanish on the lower metal plane, and has to be maximal on the opening. This creates the broad resonances in the spectra labelled by $FP_K$ in figure 9(c). These vertical Fabry-Perot resonances are governed by the semiconductor thickness $L$, as well as the metal grating thickness $h$, and can be also excited in $p$-polarization (experiment not shown). Note that they are absent for very low frequencies (the THz range), because of their finite cut-off frequency.

The horizontal contribution arises from the lateral localization between the stripes, and is governed by the parameter $a$. In order to explain this localization, once again the modal impedance mismatch between the double metal and single metal regions must be evoked. Indeed, the double metal regions of thickness $L$ support a $TE_1$ mode with a cut-off frequency of $c/2n_sL$ which is about 57 THz for the device depicted in figure 9(a), where $L=0.8\mu m$ and $n_s=3.3$ (for this wavelength range). No $o$-polarized field can therefore penetrate in the double metal region below that frequency. This condition is enforced by the requirement that $E_y$ must vanish on the vertical metal walls (thickness $h$) of the metal stripes. We can therefore assume an effective boundary condition, which requires that $E_y$ must vanish on the whole segment of length $L+h$. Taking into account both the vertical and horizontal vibrations, we can postulate a phenomenological frequency of the form:

$$\nu_{MK} = \frac{c}{2n'_M}\sqrt{\frac{M^2}{a^2} + \frac{(K-1/2)^2}{(L+h)^2}}. \qquad (34)$$

Here $n'_M$ is the effective modal index. For instance, by comparing Eq. (34) with the experimental frequency of the fundamental mode $M=1$, $K=1$ (at about 30 THz from 9(c)) we obtain $n'_M =2.3$ which is lower than the bulk index of the semiconductor ($n_s=3.3$). This can be easily understood from figure 9(c) which shows that the mode is partially localized in the air.

Like in 2.2 and 3 the selection rules for this type of resonances can be deduced by projection of the incident field on the field distribution. With the ansatz $E_y \sim \sin((M+1)\pi x/a)$ we obtain the following form factor:

$$I_M = \frac{1-(-1)^M e^{i\Phi}}{\pi^2 M^2 - \Phi^2}\pi M, \quad \Phi = \frac{2\pi a}{\lambda_{M,K}}\sin\theta. \qquad (35)$$

The function $Re(I_M)$ is rapidly decreasing with the lateral order $M$, which is consistent with the experimental spectra in fig 9(b), where the resonant absorption is rapidly extinguished for the high order resonances. Note that in this case we can no longer provide an intuitive vision for the selection rules by any induced charges and currents.

For this type of resonances, we also observe increased quality factors with respect to the localized resonances in the THz region. These larger quality factors can be easily explained by the fact that, because of the boundary conditions, the electric field $E_y$ must avoid the metal walls, and therefore the ohmic losses are greatly reduced.

## 6. Conclusion

In conclusion, we have investigated the photonic behaviour of metal-semiconductor-metal structures with an upper metallic surface structured into rectangular strip grating or grating of square patches, in the THz frequency range. The most prominent feature of these devices is their ability to support TM-polarized electromagnetic modes which are strongly localised in the double-metal regions, with resonant frequencies depending exclusively on the size $s$ of the confining regions.

We have provided an analysis of the confining mechanism, which is the impedance mismatch between the highly confined modes and the free space plane waves. This impedance mismatch is exalted when the semiconductor thickness $L$ is made very small with respect to the resonant wavelength. When the device thickness is gradually increased, the individual resonators start to couple and the electromagnetic mode with initially flat dispersion gradually degenerate into photonic crystal-like mode delocalised along the whole structure.

We have also analysed the coupling of free space photons into these highly confined sub-wavelength modes. Precise selection rules are determined based on the symmetry of the individual resonators and on the retardation effects of the incoming field. We have also shown that the absolute value of the resonantly absorbed energy is strongly impacted by the grating near field, and more precisely by the evanescent waves provided by the grating diffraction, governed by the grating period $p$. Therefore our structures are able to convert very efficiently the energy of the far field into energy of near field spatial modes. Based on recent evidence that the light-matter interaction between strongly confined modes and the electronic transitions in semiconductor quantum wells can be exalted in such structures [25], we believe that the concepts developed in this work will be useful for the development of compact subwavelength devices, such as detectors and emitters, operating not only in the THz, but also at shorter wavelengths.

**Acknowledgements**

We gratefully acknowledge support from the French National Research Agency through the programs ANR-06-NANO-047 MetalGuide, ANR-05-NANO-049 Interpol, the ERC Grant "ADEQUATE" and from the Austrian Science Fund (FWF).

**References and links**

1. W. L. Barnes, "Surface plasmon–polariton length scales: a route to sub-wavelength optics," J. Opt. A: Pure Appl. Opt. **8,** S87–S93 (2006).
2. S. A. Maier, "Plasmonic field enhancement and SERS in the effective mode volume picture," Opt. Express **14**, 1957-1964 (2006).
3. W. L. Barnes, A. Dereux, T. W. Ebbesen, "Surface plasmon subwavelength optics," Nature **424**, 824-830 (2003).
4. H. Liu and P. Lalanne, "Microscopic theory of the extraordinary optical Transmission," Nature **452**, 728-731 (2008).
5. C. Sirtori, C. Gmachl, F. Capasso, J. Faist, D. L. Sivco, A. L. Hutchinson, A.Y. Cho "Long-wavelength ($\lambda \approx$ 8–11.5 µm) semiconductor lasers with waveguides based on surface plasmons," Opt. Lett. **23**, 1366-1368 (1998).
6. R. Köhler, A. Tredicucci, F. Beltram, H. E. Beere, E. H. Linfield, A. G. Davies, D. A. Ritchie, R. C. Iotti, F. Rossi, "Terahertz semiconductor-heterostructure laser, "Nature **417**, 156-159 (2002).
7. J. B. Pendry, L. Martin-Moreno, F. J. Garcia-Vidal, "Mimicking Surface Plasmons with Structured Surfaces," Science **305**, 847-848 (2004).
8. C. R. Williams, S. R. Andrews, S. A. Maier, A. I. Fernández-Domínguez, L. Martín-Moreno, F. J. García-Vidal, "Highly confined guiding of terahertz surface plasmon polaritons on structured metal surfaces," Nature Photonics **2**, 175 -179 (2008).
9. D. R. Smith, J. B. Pendry, M. C. K. Wiltshire, "Metamaterials and Negative Refractive Index," Science **305**, 788 -792 (2004).
10. V. M. Shalaev, "Optical negative-index metamaterials," Nature Photonics **1**, 41-48 (2007).
11. M.J Adams, *An Introduction to Optical Waveguides*, (John Wiley & Sons, Chichester, 1981).
12. S. Kumar, B. S. Williams, Q. Qin, A.W. Lee, Q. Hu, J. L. Reno, "Surface-emitting distributed feedback terahertz quantum-cascade lasers in metal-metal waveguides," Opt. Express **15**, 113-128 (2007).
13. L. Mahler, A. Tredicucci, F. Beltram, C. Walther, J. Faist, H. E. Beere, D. A. Ritchie, D.S. Wiersma, "Quasi-periodic distributed feedback laser," Nat. Photonics, **4**, 165-169 (2010).

14. Y. Chassagneux, R. Colombelli, W. Maineult, S. Barbieri, H. E. Beere, D. A. Ritchie, S. P. Khanna, E.H.Linfield, A.G.Davies, "Electrically pumped photonic-crystal terahertz lasers controlled by boundary conditions," Nature **457**, 174-178 (2009).
15. Y. Chassagneux, R. Colombelli, W. Maineult, S. Barbieri, S. P. Khanna, E. H. Linfield, A. G. Davies, "Predictable surface emission patterns in terahertz photonic-crystal quantum cascade lasers," Opt. Express **17**, 9491-9502 (2009).
16. A. P. Hibbins, J. R. Sambles, C. R. Lawrence, James R. Brown, "Squeezing Millimeter Waves into Microns," Phys. Rev. Lett. **92**, 143904 (2004).
17. M. J. Lockyear, A. P. Hibbins, J. R. Sambles, P. A. Hobson, C. R. Lawrence, "Thin resonant structures for angle and polarization independent microwave absorption," Appl. Phys. Lett., **94**, 041913 (2009).
18. Y. Todorov and C. Minot, "Modal method for conical diffraction on a rectangular slit metallic grating in a multilayer structure," J. Opt. Soc. Am. A **24**, 3100-3114 (2007).
19. I. C. Botten, M. S. Craig, R. C. McPhedran, J. L. Adams, J. R. Andrewartha , "Highly conducting lamellar diffraction grating", Opt. Acta **28**, 1103-1106 (1981).
20. P. Sheng, R. S. Stepleman, P. N. Sanda "Exact eigenfunctions for square-wave gratings: Application to diffraction and surface-plasmon calculations," Phys. Rev. B **26**, 2907-2917 (1982).
21. L. Landau and E. Lifchitz, *Electrodynamics of Continuous Media* (Mir, Moscow, 1969).
22. Y. Xu, Y. Li, R. K. Lee, A. Yariv, "Scattering-theory analysis of waveguide-resonator coupling," Phys. Rev. E **62**, 7389-7404, (2000).
23. A. P. Hibbins, W. A. Murray, J. Tyler, S. Wedge, W. L. Barnes, J. R. Sambles, "Resonant absorption of electromagnetic fields by surface plasmons buried in a multilayered plasmonic nanostructure," Phys. Rev. B **74**, 073408 (2006).
24. C. Kittel, *Introduction to solid state physics* (John Wiley & Sons, 1976).
25. Y. Todorov, A. M. Andrews, I. Sagnes, R. Colombelli, P. Klang, G. Strasser, C. Sirtori, "Strong Light-Matter Coupling in Subwavelength Metal-Dielectric Microcavities at Terahertz Frequencies ," Phys. Rev. Lett., **102**, 186402 (2009).
26. *Handbook of Optical Constants of Solids*, E. Palik ed.,(Academic Press, San Diego, 1998).
27. *Electromagnetic theory of gratings*, R. Petit ed., (Springer-Verlag, Berlin, 1980).
28. C. Balanis, *Antenna Theory* (John Wiley & Sons, 2005).
29. N. H. Fletcher and T. D. Rossing, *The Physics of Musical Instruments*, (Springer-Verlag, New York, 1998).
30. P. Lalanne, J. P. Hugonin, J. C. Rogier, "Theory of Surface Plasmon Generation at Nanoslit Apertures," Phys. Rev.Lett., **95**, 263902 (2005).
31. R. Gordon, "Light in a subwavelength slit in a metal: Propagation and reflection," Phys. Rev. B. **73**, 153405 (2006).
32. As the thickness *L* is increased beyond 1µm the higher order guided modes should be taken into account in order to obtain quantitative results. Nevertheless, the qualitative behaviour described in the text remains the same.
33. The Hankel function has a non-integrable singularity at *x=0*, however this singularity can be rendered integrabale by inclusion of the metallic loss, which creates a non-zero imaginary part in the frequency $k_0$.
34. M. Bahriz, O. Crisafulli, V. Moreau, R. Colombelli, O. Painter, "Design of mid-IR and THz quantum cascade laser cavities with complete TM photonic bandgap," Opt. Express **15**, 5948-5965 (2007).
35. R. W. Wood, "Anomalous diffraction gratings," Phys. Rev. **48,** 928-937 (1935) .
36. S. Collin, F. Pardo, R. Teissier, J.-L. Pelouard, "Strong discontinuities in the complex photonic band structure of transmission metallic gratings," Phys. Rev. B. **63**, 033107 (2001).
37. L.A.Coldren and S.W. Corzine, *Diode lasers and Photonic Integrated Circuits*, (John Wiley & Sons, New York, 1995).
38. M. Sarrazin, J.-P. Vigneron, J.-M. Vigoureux, "Role of Wood anomalies in optical properties of thin metallic films with a bidimensional array of subwavelength holes," Phys. Rev. B **67**, 085415 (2003).
39. C. Cohen-Tnaoudji, B. Diu, F. Laloë, *Quantum Mechanics, Vol. I*, (Hermann, Paris, 1977).
40. Frédérique de Fornel, *Les ondes évanescentes en Optique et en Optoélectronique*, (Eyrolles, Paris 1997).
41. M. Cai, O. Painter, K. J. Vahala, "Observation of Critical Coupling in a Fiber Taper to a Silica-Microsphere Whispering-Gallery Mode System," Phys. Rev. Lett. **85**, 74-77 (2000).